\documentclass[a4paper,twocolumn,
english,aps,prl,floatfix,showpacs]{revtex4}
\usepackage{graphicx}
\usepackage{amsmath}
\usepackage{babel}
\usepackage{graphics}
\usepackage{amssymb}

\begin{document}

\title{Generalized Constraints on Quantum Amplification.}

\author{U. Gavish$^1$, B. Yurke$^2$ and Y. Imry$^3$}
\affiliation{1. LKB, Ecole Normale Superieure, Paris \\
2. Bell Labs, Lucent Technologies, Murray Hill,  NJ \\
3. Condensed Matter Physics Dept., Weizmann Institute, Rehovot}

\begin{abstract}
We derive quantum constraints on the minimal amount of noise added
in linear amplification involving input or output signals whose
component operators do not necessarily have c-number commutators,
as is the case for fermion currents. This is a generalization of
constraints derived for the amplification of bosonic fields whose
components posses c-number commutators.
\end{abstract}

\date{\today}


\maketitle

It has often been noted that Heisenberg uncertainty relations, or
the underlying operator commutation relations, impose performance
restrictions on amplifiers\cite{shimoda}-\cite{devoret set} and
detectors\cite{averin}\cite{girvin}. In particular it has been
shown\cite{caves1}\cite{caves2}\cite{Yurke denker} that a linear
phase insensitive amplifier must emit noise out of its output port
with a noise power of at least $(G^2 - 1)\hbar\omega/2$ per unit
bandwidth, where $G^2$ is the power gain.  In laser amplifiers
this noise results from spontaneous emission.  In parametric
amplifiers this noise is frequency converted and amplified Nyquist
noise from the black body absorber terminating the idler port. The
discussions have generally been carried out in the context of
amplifying bosonic fields with no particle
conservation law (simply called "bosonic" below),
such as electromagnetic fields in optical systems or collective charge
excitations in electronic circuits amplified by lasers or parametric amplifiers .
Heisenberg operators representing
such signals have the property that they can be decomposed into
two components, the commutator of which is a c-number. Therefore,
the Heisenberg principle is used to derive quantum constraints on
their amplification in a way that is similar to arguments normally
used for an ordinary pair of canonical variables. However, such an
analysis excludes a very common (perhaps even the most common)
case of a signal: one in which the commutator of the Heisenberg
operators representing the components of the input signal is an
operator instead of a c-number.

For example, in semiconductor or molecular transistors often the
signal is carried by current of fermions.  Due to the conservation
law of these particles the current operators in this case are
bilinear in the fermion creation and annihilation operators. As a
result the commutators of the current operators are themselves
operators bilinear in fermion creation and annihilation operators.
Consequently, the derivations of the quantum limits of amplifier
noise performance that have been provided for amplifiers of
 bosonic fields are
not directly applicable to amplifiers amplifying fermion currents.
To overcome this limitation we derive generalized quantum limits
of amplifier noise performance that are applicable when the input
or output fields are either fermion or boson fields.

We begin with a review of the usual analysis for  bosonic
amplifiers \cite{shimoda}-\cite{caves2}. Suppose first that the
signals can be described by a single pair of canonical variables:
Let $x$ and $p$ be the position and momentum Heisenberg operator
for the signal fed into the amplifier and let $X$ and $P$ denote
the position and momentum operators for the amplified signal.
Ideally, for a phase insensitive amplifier, one would like a
device that simply produces an amplified copy of the input:
\begin{eqnarray}
 X = G x~~~;~~~
 P = G p   \label{1}~ .
\end{eqnarray}
However, these relationships are incompatible with the
position-momentum commutation relations that must be satisfied on
each side of the amplifier:
\begin{eqnarray}\label{2}
 [x,p] = i\hbar  ~~~;~~~
[X,P] = i\hbar  ~.
\end{eqnarray}
Hence, Eqs.~(\ref{1}) and (\ref{2}) must be modified according to
\begin{eqnarray}
 X= G x + X_N ~~~;~~~
P = G p + P_N \label{3} \ .
\end{eqnarray}
where $X_N$ and $P_N$ are noise operators.  Since the noise is
uncorrelated with the signal, one has
\begin{eqnarray}
 [x, X_N] = [p, X_N]=[x, P_N] = [p, P_N] = 0 \label{4} \ .
\end{eqnarray}
Equations~(\ref{2}) through (\ref{4}) impose the following
commutation relation on the noise operators:
\begin{equation}
[P_N, X_N] = i\hbar(G^2 - 1) \label{5} \ .
\end{equation}
This immediately implies the uncertainty relation
\begin{equation}
\Delta X_N \Delta P_N \geq \frac{1}{2} \hbar{(G^2 - 1)} \
\label{6}
\end{equation}
where $\Delta A\equiv (\langle A^2 \rangle - \langle A
\rangle^2)^{1/2}.$
%

A similar argument to the above shows that a phase sensitive
amplifier performing the transformation
\begin{eqnarray}
X = G x ~~~;~~~ P = \frac{1}{G}p \label{7a}
\end{eqnarray}
is fully compatible with the commutation relations Eq.~(\ref{2}).
Such an amplifier need not add noise to the signal delivered at
the output port
\cite{caves1}\cite{caves2}\cite{yuen}\cite{hollenhorst}.

Consider now a signal carried by a current $I(t).$ Let us expand
the operator $I(t)$ in Fourier form:
\begin{eqnarray}\label{7b}
I(t)=\frac{1}{\sqrt{2\pi}}\int_{-\infty}^{\infty}d\omega
I(\omega)e^{-i\omega t}
\end{eqnarray}
For simplicity we shall consider systems small enough so that the
spatial dependence of the current may be ignored.  $I(t)$ is
Hermitian and therefore: $I(\omega)=I^{\dagger}(-\omega).$

To carry out an analysis analogous to that of Eq.~(\ref{1})
through (\ref{7a}) operators playing the role of $x$ and $p$ need
to be constructed.  To this end we introduce $I_f(t)$ which
consists of $I(t)$ passed through a square filter centered at
frequency $\pm\omega_0$ and having a width $\Delta \omega\ll
\omega_0:$
\begin{eqnarray}
I_f(t) = \frac{1}{\sqrt{2\pi}} \int_{-\omega_0 - \Delta \omega
/2}^{-\omega_0 + \Delta \omega / 2}
e^{-i\omega t} I(\omega) d\omega ~~~~~~ \nonumber\\
+\frac{1}{\sqrt{2\pi}} \int_{\omega_0 - \Delta \omega
/2}^{\omega_0 + \Delta \omega / 2} e^{-i\omega t} I(\omega)
d\omega \label{13a}
\end{eqnarray}
This can be written as:
\begin{equation}
I_f(t) = I_{1}(t)\cos(\omega_0 t) + I_{2}(t)\sin(\omega_0 t)
\label{10a}
\end{equation}
where we defined the Hermitian \emph{signal components}  by:
\begin{equation}
I_{1}(t) = \frac{1}{\sqrt{2\pi}} \int_{ - \Delta \omega
/2}^{\Delta \omega / 2} d\omega \ [ I(\omega_0 + \omega) e^{-i
\omega t} + I^\dagger(\omega_0 + \omega)e^{i \omega t}]
\label{11a}
\end{equation}
and
\begin{equation}
I_{2}(t) = \frac{-i}{\sqrt{2\pi}} \int_{ - \Delta \omega
/2}^{\Delta \omega / 2} d\omega \ [ I(\omega_0 + \omega) e^{-i
\omega t} - I^\dagger(\omega_0 + \omega)e^{i \omega t}] \ .
\label{14a}
\end{equation}
For  $I_{1}$ and $I_{2}$ to play the role
 of $x$ and $p$ above, one should obtain their commutator.
  \begin{figure}
    \begin{center}
       \includegraphics[height=3in,width=2.1in,angle=-90]{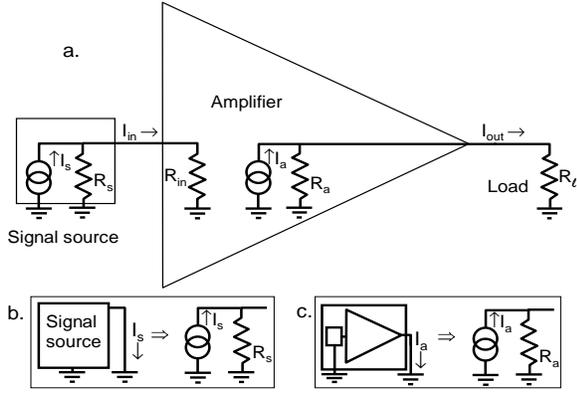}
         \caption{\textbf{a.} A source $R_s$ feeding an amplifier connected to a load $R_{\ell}.$
        \textbf{b.} The  signal source Norton equivalent.  The signal source would produces a current $I_s$ if short-circuited.
         \textbf{c.} The amplifier Norton equivalent. The signal source + amplifier would produce a current $I_a$ if short-circuited.
} \label{fig1}
     \end{center}
\vspace{-0.65cm}
\end{figure}
In bosonic systems, $I_{1}$ and $I_{2}$
are linear in bosonic creation and annihilation operators
and therefore their commutator is a c-number. For
example,  the transform of the current field propagating along a
semi-infinite ideal transmission line having impedance $R$  is
given by\cite{LouisellTransLine}\cite{Yurke TransLine}:
\begin{eqnarray} \label{Iw TransLine} I(\omega)=-i\sqrt{\frac{2\hbar\omega}{ R}}
a(\omega)\end{eqnarray} where $a(\omega),$ is the annihilation
operator of the transmission line mode at frequency $\omega.$
$a(\omega)$ satisfies
 the Bose commutation relations $[a(\omega),a^{\dagger}(\omega')]=
\delta(\omega-\omega').$
%
%
%
%
Denoting
$\Delta \nu = \frac{\Delta \omega}{2 \pi}$ and using Eqs.
(\ref{11a})-(\ref{Iw TransLine}) one can verify that the
commutator of the components of this current is indeed a c-number and is given by
\begin{equation}\label{commut TransLine}
[I_{1}(t),I_{2}(t)]=i4\hbar\omega_0\frac{1}{R}\Delta \nu.
\end{equation}

Consider now an amplifier connected to a signal source at its
input port and to a load resistor at its output port (Fig. 1.a).
To avoid the specific details of the source, it is modelled by its
quantum Norton equivalent circuit \cite{Norton}, i.e., by a
resistor $R_s$ in parallel to a current source $I_s(t)$ where
$R_s$ is the total resistance of the signal source between its two
terminals and $I_s(t)$ is the \emph{short-circuited current
operator}, i.e., it is the Heisenberg current operator of the
signal source if the amplifier is replaced by a short (Fig. 1.b).
When coupled to an amplifier having an impedance $R_{in}$ at its
input port, the source  $I_s(t)$ delivers a current $I_{in}(t)$
into it. If $R_{in}$ is not a short than only a part of $I_s(t),$
namely  $I_{in}(t)=I_s(t)R_s/(R_s+R_{in}),$ will be delivered into
the amplifier. Similarly,  the combined system of the signal
source and the amplifier acts as a resistance $R_a$ in parallel to
a current source $I_a(t)$ (Fig. 1.c) which delivers a current
$I_{out}(t)=I_a(t)R_a/(R_a+R_{\ell})$ to a load resistor
$R_{\ell}.$ In analogy with Eq.~ (\ref{3}) we write:
\begin{equation}
I_{out}(t) = G \left(\frac{R_s}{R_{\ell}}\right)^{1/2} I_{in}(t) +
I_{N}(t)\label{18}
\end{equation}
where $G^2$ is the power gain and $I_N(t)$ is a noise operator
which commutes with $I_{in}(t)$. The currents $I_{out}(t)$ and
$I_{in}(t)$ are observable and, hence Hermitian. As a consequence
$I_N(t)$ is also Hermitian. Written in terms of the signal
components Eq.~(\ref{18}) becomes:
\begin{eqnarray}
 I_{out,q}(t) = G \left(\frac{R_s}{R_{\ell}}\right)^{1/2} I_{in,q}(t) +
I_{Nq}(t)~~~q=1,2. \label{19a}
\end{eqnarray}
As in Eq.~(\ref{4})  the noise is taken to be uncorrelated with
the signal.  Consequently:
\begin{equation}
[I_{in,\beta}(t), I_{N\beta'}(t)]  = 0 ~~~~~~\beta,\beta'=1,2 \ .
\label{40c}
\end{equation}
Assuming $I_s$ and $I_a$ obey the bosonic relation, Eq.
(\ref{commut TransLine}), with $R$ replaced by $R_s$ and $R_a$
respectively, one has
\begin{equation}
[I_{\beta1}(t), I_{\beta 2}(t)] = i 4 \hbar\omega_0
\frac{1}{R_{\beta}} \Delta \nu ~~~~~~\beta=a,s~. \label{23x}
\end{equation}
We shall restrict ourselves to determining the optimum performance
of power amplification when
 maximum power is transferred from the signal source to the amplifier and
 from the amplifier to the load. In this case the amplifier should be impedance-matched to the source and the load and therefore one has
$R_s=R_{in}$ and $R_a=R_{\ell}.$ This implies that the current
sources $I_s$ and $I_a$ deliver only half of their current to the
amplifier input and the load, i.e., $I_s=2I_{in}$ and
$I_a=2I_{out}.$ Thus, Eq. (\ref{23x}) yields
\begin{eqnarray}
[I_{in,1}(t), I_{in,2}(t)] = i  \hbar\omega_0\frac{1}{R_s}
\Delta \nu ~~ \nonumber \\
~[I_{out,1}(t), I_{out,2}(t)] = i  \hbar\omega_0\frac{1}{R_{\ell}}
\Delta \nu ~.
 \label{24x}
\end{eqnarray}
Replacing $x,~p,~X$ and $P$ by $I_{in,1},~I_{in,2},~I_{out, 1}$
and $I_{out, 2},$ using the commutator Eq. (\ref{24x}) instead of
Eq. (\ref{2})
  and following the steps leading to Eq.
(\ref{6}) one obtains the well-known constraint on the minimal
noise that the bosonic amplifier
 must add to the signal\cite{caves1}\cite{caves2}\cite{Yurke denker}:
\begin{equation}\label{25x}
\Delta I_{N1}(t)\Delta I_{N2}(t) \geq
(G^2-1)\frac{\hbar\omega_0}{2}\frac{1}{R_{\ell}}\Delta \nu~.
\end{equation}

In the derivation of Eq. (\ref{25x}) an essential use was made of
the fact that the commutation relation of the bosonic currents,
Eq. (\ref{24x}), was a c-number as in Eq. (\ref{2}). However, as
explained above, for other types of signals this is not the case.
That is, one generally has
\begin{equation}\label{commut TransLine general}
[I_{1}(t),I_{2}(t)]\neq c~number \ ,
\end{equation}
in which case the derivation given above is not justified.

We shall now derive generalized constraints on amplifier
noise performance which will be valid regardless of whether the
commutators of the components of the current operator are
c-numbers or operators. In particular, these constraints are valid for both
fermionic and bosonic currents. We assume that the amplifier and
the signal are both in stationary states and again take the
input-output relations, Eq. (\ref{19a}),
 except that the \emph{differential
conductances} (see below) of the amplifier and the source, $g_{s}$
and $g_{\ell},$ replace the ordinary ones, $R_s^{-1}$ and
$R_{\ell}^{-1}:$
\begin{equation}
I_{out, q}(t) = G \left(\frac{g_{\ell}}{g_s}\right)^{1/2}
I_{in,q}(t) + I_{N\beta}(t) ~~~~q = 1,2~.\label{25y}
\end{equation}
Since the signals under consideration include those with the
property Eq. (\ref{commut TransLine general}), we do not have a
simple commutator to work with.  However, Kubo's
fluctuation-dissipation theorem, generalized to nonequilibrium
steady states, provides a means of carrying out the analysis by
supplying a substitute for such a commutator.  This is the key
element in our derivation of the generalized constraints.
 Denoting
\begin{equation}
S(\omega) = \int_{-\infty}^\infty dt e^{i\omega t} \langle I(0)
I(t) \rangle \ , \label{26y}
\end{equation}
the Kubo theorem states that \cite{kubo0}-\cite{moriond 2001}
\begin{equation}
S(-\omega) - S(\omega)= 2\hbar \omega g(\omega)~~~~~\omega >0
\label{27y}
\end{equation}
where $g(\omega)$ is the differential conductance of the system at frequency
$\omega.$ Simple derivations of Eq. (\ref{27y}) are given in Refs.
\cite{Canary},\cite{moriond 2001}. An important aspect of
Eq.(\ref{27y}) is that it holds regardless of whether the current
is carried by bosons or fermions. Another important aspect of it
is that it holds for \emph{any} stationary state including
nonequilibrium ones\cite{non stat}. The current appearing in Eq. (\ref{26y}) may
have a non-zero expectation value, $\langle I(t)\rangle \neq 0,$
since the system is not necessarily in equilibrium. If one applies
a small AC voltage, $V_{AC}=Ve^{i\omega t}$ (on top of any other
field that may already exist and is driving the system out of
equilibrium), then an additional current proportional to $V_{AC}$
appears: $\langle I_V(t)\rangle =\tilde{ \sigma}(\omega)
Ve^{i\omega t}.$ The differential conductance is defined as the
real part of the proportionality coefficient,
$\tilde{\sigma}(\omega),$ $g (\omega)\equiv Re
\tilde{\sigma}(\omega).$

We now derive some  expressions that follow from Eqs.~(\ref{26y})
and (\ref{27y}).  First, we combine them into:
\begin{equation}
\int_{-\infty}^\infty d\tau e^{i\omega \tau} \langle [I(\tau),
I(0)] \rangle = 2 \hbar \omega g(\omega) \ . \label{28y}
\end{equation}
Recalling the assumption of stationarity, one can show that
Eq.~(\ref{26y}) implies:
\begin{eqnarray}
 \langle I(\omega)I(\omega') \rangle
= \delta(\omega + \omega')S(-\omega) \label{23} ~~~~~~~~~~~~~\\
\langle I(\omega)I^\dagger(\omega') \rangle
= \delta(\omega - \omega')S(-\omega) \label{24} ~~~~~~~~~~~~\\
 \langle I^\dagger(\omega)I(\omega') \rangle
= \delta(\omega - \omega')S(\omega) \label{25} ~~~~~~~~~~~~~~\\
\langle I^\dagger(\omega)I^\dagger(\omega') \rangle
= \delta(\omega + \omega')S(\omega) \label{26} ~~~~~~~~~~~~~\\
\langle [I(\omega),I(\omega')] \rangle = \langle
[I^{\dagger}(\omega),I^{\dagger}(\omega')] \rangle = 0,
~~~\omega,\omega' >0. \label{29y}
\end{eqnarray}
From these relationships and Eq.~(\ref{27y}) one also has
\begin{equation}
\langle [I(\omega),I^\dagger(\omega')] \rangle = 2 \delta(\omega -
\omega') \hbar \omega g(\omega) \ . \label{30y}
\end{equation}
From Eqs. (\ref{11a}), (\ref{14a}), (\ref{29y}), and (\ref{30y})
one obtains (compare with Eq. (\ref{commut TransLine})):
\begin{equation}
\langle [I_{1}(t), I_{2}(t)] \rangle = 4i\hbar \omega_0 g \Delta
\nu  \ . \label{31y}
\end{equation}
where we assumed for simplicity that $g$ is independent of the
frequency. The generalization to the frequency-dependent case is
straightforward.

To be able to carry out an amplifier noise analysis we first have
to identify the relation between the currents appearing in Eqs.
(\ref{25y}) and (\ref{31y}). As in Eqs. (\ref{26y}) and
(\ref{27y}), the current and the conductance appearing in Eq.
(\ref{31y}) are the Heisenberg current operators and the
differential conductance of the system obtained when the voltage,
or chemical potential is not allowed to fluctuate (as is the case
when a 2-terminal device is connected across a zero impedance
device such as a short or a voltage source). By their definition
these are just, respectively,  the currents produced by the
current sources and the differential conductances of the resistors
in the Norton circuits described above Eq. (\ref{18}). Therefore,
Eq. (\ref{31y}) applies for $I_s$ and $I_a$:
\begin{equation}
\langle [I_{\beta 1}(t), I_{\beta 2}(t)] \rangle = 4i\hbar
\omega_0 g_{\beta} \Delta \nu  ~~~~~~\beta=s,a~ . \label{32y}
\end{equation}
As above, assuming the impedance-matching conditions, $g_s=g_{in}$
and $g_a=g_{\ell}$, imply that the currents delivered to the
amplifier and to the load are only half those of
 the Norton equivalent current generators, hence, $I_s(t)=2I_{in}(t),~I_a(t)=2I_{out}(t).$
Eq. (\ref{32y}) now becomes:
\begin{eqnarray}
\langle [I_{in,1}(t), I_{in,2}(t)] \rangle = i\hbar \omega_0 g_s
\Delta
\nu  ~~\nonumber \\
\langle [I_{out,1}(t), I_{out,2}(t)] \rangle = i\hbar \omega_0
g_\ell \Delta \nu   ~ . \label{33y}
\end{eqnarray}
Eqs.~(\ref{40c}), (\ref{25y}) and (\ref{33y}) yield:
\begin{equation}
\langle [ I_{N2}(t),I_{N1}(t) ] \rangle = i(G^2 - 1) \hbar
\omega_0 g_{\ell}\Delta \nu \ . \label{35y}
\end{equation}
For the next step of the derivation we recall that for any two
Hermitian operators $A$ and $B$ one has
\begin{equation}
\Delta A \Delta B \geq  \frac{1}{2}|\langle [A,B] \rangle | ,
\label{36y}
\end{equation}
where $[A,B]$ may be either an operator or a c-number. It follows
that
\begin{equation}
\Delta I_{N2} \Delta I_{N1} \geq (G^2-
1)\frac{\hbar\omega_0}{2}g_{\ell} \Delta \nu \ . \label{37y}
\end{equation}
Eq. (\ref{37y}) is our main result. It is valid regardless of
whether the signal is carried by fermions or bosons. In
particular, it does not require the commutator of the current
components to be a c-number,  since only the expectation value of
such a commutator, Eq. (\ref{31y}), enters into the derivation.
For specific fermionic devices, such as mesoscopic or molecular
amplifiers, Eqs.~(\ref{35y}) and (\ref{37y}) allow one to obtain
constraints on device parameters that must be met in order for
the device to achieve quantum limited noise performance
\cite{molec amplif moriond 2004}.

The expectation value of the time averaged noise power delivered
to the load in the frequency band of width $\Delta \omega$ about
$\omega_0$ is given by
\begin{equation}
P_{\ell N} = \frac{1}{g_{\ell}} \langle I_{N}^2(t) \rangle =
\frac{1}{2g_{\ell}}[\langle I_{N1}^2\rangle + \langle I_{N2}^2
\rangle ] . \label{38y}
\end{equation}
Taking the noise to have zero mean and the the amplifier to be
truly phase insensitive so that $\Delta I_{N2} = \Delta I_{N1},$
Eq.~(\ref{37y}) yields
\begin{equation}
P_{\ell N} \geq \frac{\hbar \omega_0}{2} (G^2 - 1) \Delta \nu \ .
\label{39y}
\end{equation}

Using Eq.~(\ref{32y}) it is also straightforward to demonstrate
that it is permissible for an amplifier to noiselessly carry out
the transformation:
\begin{eqnarray}
I_{out, 1}(t) = G \sqrt{\frac{g_{\ell}}{g_s}} I_{in,1}(t)\nonumber \\
 I_{out, 2}(t) = \frac{1}{G} \sqrt{\frac{g_{\ell}}{g_s}}
I_{in,2}(t) \label{40y}  .
\end{eqnarray}

In conclusion, using the nonequilibrium Kubo formula, one can
generalize the usual noise limits of amplifiers to the case in
which the commutator of the signal components is not a c-number
as, e.g., is the case of a current carried by fermions. Again, for
phase insensitive amplifiers, a noise power of $(G^2 - 1)\hbar
\omega/2 $ per unit bandwidth must be added to the amplified
signal.  Furthermore, this analysis allows for the possibility of
phase sensitive fermion based devices which contributes no noise
to the component of the signal that is amplified.

U. G. and Y. I. are thankful for illuminating discussions
with Y. Levinson and B. Dou\c{c}ot. Research at WIS was supported by a Center of
Excellence of the Israel Science Foundation (ISF) and by the
German Federal Ministry of Education and Research (BMBF), within
the framework of the German Israeli Project Cooperation (DIP).


\begin{thebibliography}{99}
\bibitem{Norton}
The Norton equivalent is a representation of an output port of
physical system or device that could be arbitrarily complex inside
and could contain power sources such as batteries.  The system
could thus be out of equilibrium, in which case the current source
of the Norton equivalent need not have equilibrium noise
statistics. See e.g., and Sec. V in \cite{Yurke denker}.
\end{thebibliography}
\end{document}